\documentclass
[12pt]{article}
\usepackage[dvips]{color}
\usepackage{epsfig}
\usepackage{amsmath}
\usepackage{graphicx}
\begin{document}
\title {The Holographic Description of Baryon in Non- Critical String Theory }
\author{M. R. Pahlavani$^{a}$\thanks{Email:
m.pahlavani@umz.ac.ir.ir}\hspace{1mm}, J. Sadeghi $^{a}$\thanks{Email:
pouriya@ipm.ir}\hspace{1mm} and  R. Morad$^{a}$\thanks{Email:
r.morad@umz.ac.ir}\hspace{1mm}\\
$^a$ {\small {\em  Sciences Faculty, Department of Physics, Mazandaran University,}}\\
{\small {\em P. O. Box 47415-416, Babolsar, Iran}}}
 \maketitle
\begin{abstract}
We consider a holographic model constructed from an intersecting brane configuration $D4/D4/\overline{D4}$
in noncritical string theory. We study the baryon in the confined phase of this supergravity by considering the
source term for the baryon. Also, the thermodynamics functions are studied. Moreover, we obtain the binding
energy of the baryon in this holographic QCD-like model.
\\
{\bf Keywords}: AdS/CFT correspondence; Noncritical string theory; Baryon.
\end{abstract}
\newpage
\section{Introduction}
The AdS/CFT correspondence is a useful duality between string
theory in $d+1$ dimensional space-time and conformal field theory
in $d$ dimension [1-4]. It also can be expanded to the general
cases of string-gauge dualities like non conformal and non
supersymmetric theories. So, some more realistic effective
theories can be constructed from the string theory. In the low
energy physics, in case of effective boundary theories, it can be
investigated by their classical supergravity duals. Recently,
some holographic models are introduced via the gauge-gravity
correspondence that studied some features of the low-energy
QCD [7-18]. These models, called holographic QCD models, are
constructed from the intersecting brane configuration, like the
Sakai-Sugimoto model[10,11] that is a $D4/D8/\overline{D8}$.
\\In the holographic model arising from the critical string theory, the color brane backgrounds are ten-dimensional,
so the dual gauge theories are supersymmetric. In order to break
the supersymmetry, some parts of such backgrounds need to be
compactified on some manifolds. This causes to produce some
Kluza-Klein modes. The mass scale of these modes are at the same
order as the masses of the hadronic modes. These modes are
coupled to the hadronic modes. There is no mechanism
to disentangle these unwanted modes from the hadronic modes yet. One
can consider the color brane configuration in the noncritical
string theory to overcome this problem. The result is a
gravitational backgrounds located at the low dimensions. In these
backgrounds the string coupling constants are proportional to
$\frac{1}{N_c}$, so the large $N_c$ limit corresponds to the
small string coupling constant. However, contrary to the critical
holographic models, in large $N_c$ limit the 't Hooft coupling
is of order one instead of infinity and the scalar curvature of the
gravitational background is  also of order one[19,21]. So the
noncritical gauge-gravity correspondence is not very reliable.
But studies show that the results of these models for the some
low energy QCD properties like the meson mass spectrum, wilson
loop and the mass spectrum of glueballs [22-24] are comparable
with the lattice computations. Therefore, the noncritical
holographic models seem still useful to study the QCD stuff.
\\One of the noncritical holographic models is composed of $D4$ and anti $D4$ brane in six-dimensional
 noncritical string theory[20,23]. The low energy effective theory on the intersecting brane configuration is a
 four-dimensional QCD-like effective theory with the global chiral symmetry $U(N_f)_L \times U(N_f)_R$.
 In this brane configuration, the six-dimensional gravity background is the
 near horizon geometry of the color $D4$ brane. Here, there is no compact sphere compare with the critical
 gravity backgrounds. This model is based on the compactified $AdS_6$ space-time with a constant dilaton.
 So the model dose not suffer from the large string coupling as the SS model.
 The meson spectrum [22] and the structure of thermal phase [25] are studied in this model.
 Some properties, like the dependence of the meson masses on the stringy mass of the quarks and the excitation number
 are different from the critical holographic models such as SS model.
\\Baryon was analyzed in the critical holographic models like the SS model[26-43]. In the SS model, a $D4$ brane
wrapped on a $S^4$ is the baryon vertex. $N_c$ fundamental
strings attach the vertex to the flavor $D8$ brane. It was shown
that this baryon corresponds to an instanton of the
five-dimensional effective $U(N_f)$ gauge theory[43]. The
physical properties of this baryon like the mass, size, mass
splitting, the mean radii, magnetic moments and various couplings
were analyzed in the SS model[31-42]. The obtained results show a
better agreement with the experimental data compared to the
Skyrme model results[44]. But there are some problems. For
example, the size of the baryon is proportional to
$\lambda^{-1/2}$. In the large 't Hooft coupling (large
$\lambda$) the size of the baryon becomes zero. So the stringy
corrections have to be taken into account. Another problem is
that the scale of the system associated with the baryonic
structure is roughly half the one needed to fit to the mesonic
data [42]. So, all above information give us motivation to analyze
the baryon in a noncritical holographic model with the $AdS_6$
background. In this background there is no compact $S^4$ sphere,
so we consider an unwrapped $D0$ brane as a baryon vertex. Similar to the
SS model, in this case also it is necessary that $N_c$ fundamental
strings attach the vertex to the probe flavor brane. Also the
baryon vertex is attached to the color probe brane[43]. In this
paper, using this description of the baryon we calculate the
baryon energy. Also we study the thermodynamics functions
and the binding energy of baryon. We compare our results with the
results of the critical holographic model such as SS model, and
show that the behavior of the thermodynamics function respect to the baryon density is similar to the SS model [33].
Also, we find a similar behavior for the baryon binding energy, unless the value of the binding energy is larger as
compare with the SS model [14].
\section{\textbf{ $AdS_6$} Backgrounds}
In analogy with the critical models, in this noncritical model, the gravity background is
generated by near-extremal $D4$ branes wrapped over a circle with
the anti-periodic boundary conditions. Two stacks of flavor
branes, that is one of $D4$ branes and the other one of anti-$D4$
branes are added to this geometry which called flavor probe
branes. The color branes extend along the directions
$t,x_1,x_2,x_3,\tau$ while the probe flavor branes fill the whole
 Minkowski space and stretch along the radius $U$ up to
infinity. The strings attached color D4-brane to a flavor brane
transform as quarks, while strings hanging between a color $D4$ and
a flavor $\overline{D4}$ transform as anti-quarks. The chiral symmetry
breaking is achieved by a reconnection of the brane-anti-brane
pairs. Under the quenched approximation $(N_c \gg N_f)$, the
backreactions of flavor branes on the color branes can be
neglected. Just like the SS model, the $\tau$ coordinate is
wrapped on a circle and the anti-periodic condition is considered
for the fermions on the thermal circle. The final low energy
effective theory on the background is a four-dimensional QCD-like
effective theory with the global chiral symmetry $U(N_f)_L\times
U(N_f)_R$.
\\The near horizon gravity background at low energy is [20,23]
\begin{eqnarray}
ds^2&=&\left( \frac{U}{R} \right)^2 (-dt^2+dx_i dx_i+f(U) d\tau^2)
+\left( \frac{R}{U} \right)^2 \frac{dU^2}{f(U)}
\end{eqnarray}
The background geometry consists of a constant dilaton and a RR
six-form field strength as follows,
\begin{eqnarray}
F_{(6)}&=&Q_c \left( \frac{U}{R} \right)^4 dt \wedge
dx_1\wedge dx_2 \wedge dx_3  \wedge du \wedge d\tau\\
e^\phi &= &\frac{2\sqrt2}{\sqrt3 Q_c}\,\,,
\quad\qquad
\end{eqnarray}
and
\begin{eqnarray}
f(U)=1-\left( \frac{U_{KK}}{U} \right)^5,\quad\qquad \quad\qquad R^2=\frac{15}{2},
\end{eqnarray}
where  $Q_c$ is proportional to the number of color branes $N_c$.
\\To avoid the singularity, the coordinate $\tau$ satisfies the following periodic condition,
\begin{eqnarray}
\tau \sim \tau + \delta \tau\,\,,\qquad\qquad
\delta\tau=\frac{4\pi R^2}{5 U_{KK}}\,\,.
\end{eqnarray}
The Kluza-Klein mass scale of this compact dimension is
\begin{eqnarray}
M_{KK}=\frac{2\pi}{\delta\tau}=
\frac{5}{2} \frac{ \ U_{KK}}{R^2}.
\end{eqnarray}
and  dual gauge field theory in this background is non supersymmetric.
\section{Baryon in the Noncritical Holographic Model}
In this section, we consider the baryon in the noncritical
holographic model with the $AdS_6$ background. In the critical
holographic model like the SS model, $D4$ brane wrapping the
compact $S^4$ is introduced as a baryon vertex which has $N_c$
units of electric charge. It is shown that a $D4$ brane wrapping
$S^4$ looks like an object with electric charge with respect to
the gauge field on $D8$ and it is possible to say that $D4$ brane
spread inside $D8$ brane as an instanton.
\\Here we consider noncritical holographic model which has no compact sphere. So one can introduce an
unwrapped $D0$ brane as a baryon vertex. There is a Chern-Simon
term on the $D0$ brane world-volume. Also in this case one needs
to attach $N_c$ strings to the baryon vertex. The other end the
corresponding  strings must be attached to the probe flavor $D4$
branes. Also the baryon vertex will be attached to the probe
branes [43]. Now, we turn on only the zero component of the gauge
field on the world-volume of the flavor $D4$ brane and assume that
this component depends only on the compact coordinate $\tau$.
Therefore, the abelaian effective action on the $D4$ brane is
written  by,
\begin{equation}
S_{\textrm{D4}} =
 -N_f T_4 e^{-\phi} \int d^5x
  \sqrt{-det(g_{MN}+2\pi \acute{\alpha}F_{MN})} +\mu \int C_5 ,
\end{equation}
where $T_4=(2\pi)^{-4}(l_s)^{-5}$ is the tension of $D4$ brane, the $F_{MN}=\partial_M\, A_N-\partial_N\, A_M-i[A_M,A_N],\,(M,N=0,1,..5)$ is the field strength tensor and the $A_M$ is the $U(N_f)$ gauge field on the $D4$ brane. The second term in the above action, is the Chern-Simons action which has to be zero[25]. So, we can neglect this term.
\\The induced metric on the $D4$ brane is written as,
\begin{eqnarray}
ds^2&=&\left( \frac{U}{R} \right)^2 (\eta_{\mu \nu}dx^{\mu} dx^{\nu})+
[\left( \frac{R}{U} \right)^2\,f(U)\acute{\tau}^2+\left( \frac{R}{U} \right)^2\,f(U)^{-1}]dU^2.
\end{eqnarray}
Thus, the $D4$ brane action has the following form,
\begin{equation}
S_{\textrm{D4}} =
 -N_f T_4 e^{-\phi} \int d^4x
\, dU (\frac{U}{R})^5 \sqrt{\acute{\tau}^2\,f(U)+(\frac{R}{U})^4(f(U)^{-1}-(2\pi \acute{\alpha} \acute{A_0}^2)},
\end{equation}
where the $\acute{A_0}$ and $\acute{\tau}$ are derivatives respect to the $U$ coordinate.
The equations of motion for $\tau$ and $A_0(U)$ are,
\begin{eqnarray}
\frac{d}{dU}(\frac{\acute{\tau} \, f(U)}{\sqrt{\acute{\tau}^2\,f(U)+(\frac{R}{U})^4(f(U)^{-1}-(2\pi
\acute{\alpha} \acute{A_0}^2)}}),
\nonumber\\
\frac{d}{dU}(\frac{\frac{U}{R} \,\acute{A_0} }{\sqrt{\acute{\tau}^2\,f(U)+(\frac{R}{U})^4(f(U)^{-1}-(2\pi
\acute{\alpha} \acute{A_0}^2)}}),
\end{eqnarray}
respectively.
\\For convenience, we assume that $D4$ and $\overline{D4}$
branes have the maximum separation at the boundary. In analogy
with the SS model[45], it implies that in the confined phase the
$D4$ and $\overline{D4}$ branes connect together at the
$U=U_{KK}$. In this case, we have  $\acute{\tau}=0$. We use this
simplified condition for the $\tau$.
\\By introducing the following variable,
\begin{eqnarray}
U=(U_{KK}^5+U_{KK}^3\,z^2)^{1/5},
\end{eqnarray}
and using dimensionless parameters,
\begin{eqnarray}
Z=\frac{z}{U_{KK}},\,\,K(Z)=1+Z^2.
\end{eqnarray}
the $D4$ brane action rewritten as,
\begin{equation}
S_{\textrm{D4}} =-B \int d^4x \, dZ K^{3/10} \sqrt{1-\acute{B}K^{3/5} (\partial_Z A_0)^2}
\end{equation}
where the constants $B$ and $\acute{B}$ are,
\begin{eqnarray}
B=\frac{2}{5}\frac{U_{KK}^4\,N_f T_4 e^{-\phi}}{R^3}\, ,\,\,\,\,\acute{B}=(\frac{5\,\pi \, \acute{\alpha}}{U_{KK}})^2
\end{eqnarray}
Now, in analogy with the SS model, we introduce the source term
for the baryon. This term arises from the coupling between the
gauge field $A_0$ on the flavor brane and the $N_c$ units of
electric charge on the baryon vertex. We assume that the baryon
is distributed homogenously in the $R^3$ space. Therefor, we consider
the source term as follows,
\begin{eqnarray}
S_{\mathrm{source}}= N_c n_B \int d^4x \int dZ\, \delta
(Z)\mbox{${\cal A}$}_0(Z) .
\end{eqnarray}
We assume the baryon action as a  sum of the DBI action for the
$D4$ brane and the source term. Thus, the Lagrangian density for
the baryon can be written as,
\begin{equation}
\mathcal{L}_{Baryon} =
 -B \, K^{3/10}\, \sqrt{1-\acute{B}K^{3/5} (\partial_Z A_0)^2}+\,N_c n_B \, \delta
(Z)\mbox{${\cal A}$}_0(Z)
\end{equation}
So, the equation of motion for the gauge field becomes,
\begin{eqnarray}
\frac{d}{dZ}\frac{\partial\,\mathcal{L} }{\partial(\partial_ZA_0)} \,=\,\frac{1}{2}n_q\,\delta(Z).
\end{eqnarray}
where $n_q=n_B\,N_c$ is the quark density. The conjugate momentum of the gauge field can be defined as follows,
\begin{eqnarray}
D\,\equiv\, \frac{\partial\,\mathcal{L} }{\partial(\partial_ZA_0)}.
\end{eqnarray}
So, from the equation of motion we obtain,
\begin{eqnarray}
D\,=\, \frac{1}{2}n_q\,Sgn(Z),
\end{eqnarray}
where $Sgn(Z)$ is the sign function and is determined by the symmetry between $D4$ and $\overline{D4}$.
Integrating relation (18) yields the classical solution for the gauge field,
\begin{eqnarray}
A_0(Z;n_q) = A_0(0)+ \int_{0}^{Z}\,dZ \frac{n_q/2}{\sqrt{{(B\,\acute{B})^2K^{9/5}}
+ \acute{B}\, n_q^2/4\,K^{3/5} }}\ .
\end{eqnarray}
The ``baryon charge chemical potential of a
quark'', $\mu$ is defined the boundary value of the gauge field [33,46],
\begin{eqnarray}
\mu(n_q) \equiv \lim_{|Z|\longrightarrow \infty} A_0(Z;n_q)  \ .
\end{eqnarray}
Moreover the chemical potential of the baryon is defined by,
\begin{eqnarray}
\mu_B = m_B + N_c \mu \ .
\end{eqnarray}
where the $m_B$ is the rest mass of the baryon. The variation of the $A_0(Z)$ as a function of $Z$ coordinate
has been shown in Figure 1. Also, Figure 2 shows variation of chemical potential, $\mu$ as a function of
baryon densities. In all figures $n_B$ is normalized to $(n_B/n_0)$, with the nuclear matter
density,$n_0 = 0.17 fm^{-3} \simeq 1.3\times 10^6 MeV^3.$ Also we use the $N_c=3$, $N_f=2$, $M_{KK}=1GeV$
and $\acute{\alpha}=1$ in our calculations. Figure 1 indicates that the gauge field in the large $Z$ becomes
constant for each value of $n_B/n_0$.
\\Now, we try to eliminate the gauge field in the action. In order to do
this, a Legendre transformation for the Lagrangian
density must be applied as follows,
\begin{eqnarray}
\mathcal{L}\rightarrow -\mathcal{L}_{Baryon}+D\, \acute{A_0} .
\end{eqnarray}
where $D$ is the conjugate momentum of gauge field which is defined by equations (18), (19).
It should be noted that the Legendre transformation at the classical field theory of bulk can be described as a
Legendre transformation between the canonical and grand canonical ensembles at the boundary thermodynamics[46].
\\ $A_0(Z)$ is an auxiliary field with no time dependence, so we can eliminate it by the equation (20).
So, we obtain the energy $U(n_q)$ as follows,
\begin{eqnarray}
U (n_q) &=& \int dx^3\int_{-\infty}^{+\infty} dZ\, ( - \mathcal{L})
\nonumber\\
&=&  B V \int_{-\infty}^{+\infty} dZ\,  K^{3/10}\sqrt{1 +
\frac{n_q^2}{ 4 B^2 \acute{B}}K^{-6/5}} ,
\end{eqnarray}
where $V$ represent the integral of the space part. It is known that the chemical potential is related to free
Helmholtz energy by the Gibbs equation $\mu=\frac{\partial F(n_q)}{\partial n_q}$. The free Helmholtz energy is
the internal energy $U(n_q)$ at the zero temperature. So, the chemical potential can be obtained using the Gibbs
equation and the equation (24) as follows,
\begin{eqnarray}
\mu =  \int_{-\infty}^{\infty}\,dZ \frac{n_q/4}{\sqrt{{(B\,\acute{B})^2K^{9/5}}
+ \acute{B}\, n_q^2/4\,K^{3/5} }}\ .
\end{eqnarray}
which is expected from the equations (20) and (21). In fact, $\mu$ is the work required to bring a charge from the UV
to the IR region against electric field which is same as the work done to add a quark to the system.
\\The free Helmholtz energy is given by the Gibbs relation,
\begin{eqnarray}
\frac{F}{V}= B\int_{-\infty}^{+\infty} dZ\,  K^{3/10}(\sqrt{1 +
\frac{n_q^2}{ 4 B^2 \acute{B}}K^{-6/5}}-1)
\end{eqnarray}
For convenience, we normalize the potentials with the $V$.
We can obtain the grand canonical potential by evaluating the on-shell action,
\begin{eqnarray}
\frac{\Omega}{V}= B\int_{-\infty}^{+\infty} dZ\,  K^{3/10}(\frac{1}{\sqrt{1 +
\frac{n_q^2}{ 4 B^2 \acute{B}}K^{-6/5}}}-1)
\end{eqnarray}
Therefore, the pressure is obtained as,
\begin{eqnarray}
P&=&-\frac{\Omega}{V}
\nonumber\\
&=& B\int_{-\infty}^{+\infty} dZ\,  K^{3/10}(1-\frac{1}{\sqrt{1 +
\frac{n_q^2}{ 4 B^2 \acute{B}}K^{-6/5}}})
\end{eqnarray}
and $\tilde{\mu}_B\equiv \mu_B-m_B$ is,
\begin{eqnarray}
\tilde{\mu}_B=  N_c\, \int_{-\infty}^{\infty}\,dZ \frac{n_q/4}{\sqrt{{(B\,\acute{B})^2K^{9/5}}
+ \acute{B}\, n_q^2/4\,K^{3/5} }}\ .
\end{eqnarray}
The thermodynamics functions versus the $n_B/n_0$ have been shown in figure 3.
\\Also, In the small $(n_B/n_0)$ limit, we can simplify the thermodynamics functions as follows,
\begin{eqnarray}
\mu &=& \frac{N_c\,n_0\,n}{2\,B\,\acute{B}} \int_{0}^{\infty}\,dZ K^{-9/10}\,
(1-\frac{(N_c\,n_0\,n)^2}{8\,B^2\,\acute{B}}K^{-6/5}+...)
\nonumber\\
&=&\frac{0.919\,N_c\,n_0\,n}{B\,\acute{B}}-\frac{0.047\,(N_c\,n_0\,n)^3}{B^3\,\acute{B}^2}
\end{eqnarray}
\begin{eqnarray}
U(n_B) = \frac{(N_c\,n_0\,n)^2}{4\,B\,\acute{B}} \int_{0}^{\infty}\,dZ K^{-9/10}\,
=\frac{1.84\,(N_c\,n_0\,n)^2}{4\,B\,\acute{B}}
\end{eqnarray}
\begin{eqnarray}
\Omega &=&- \frac{(N_c\,n_0\,n)^2}{4\,B\,\acute{B}} \int_{0}^{\infty}\,dZ K^{-9/10}=
-\frac{1.84\,(N_c\,n_0\,n)^2}{4\,B\,\acute{B}}
\end{eqnarray}
\begin{eqnarray}
P &=& \frac{(N_c\,n_0\,n)^2}{4\,B\,\acute{B}} \int_{0}^{\infty}\,dZ K^{-9/10}=
\frac{1.84\,(N_c\,n_0\,n)^2}{4\,B\,\acute{B}}
\end{eqnarray}
\\At low densities, internal energy, pressure and grand potential are quadratic in $(n_B/n_0)$.
But the behavior of chemical potential is different. Generally, the thermodynamics functions behave
same as the SS model in the small baryon densities. This behavior well explained in ref[33].
The small density limit can be interpreted that in bulk the $A_0$ configuration for fixed charge
is obtained by minimizing the induced DBI action on $D4-\overline{D4}$. So, there is only the
flavor meson mediated interactions between the point-like baryons. At the large $N_c$ limit $D4$
mediated correlated gravitons are heavy and decouple, so the point-like baryon vertex on the bulk can
be considered as the skyrmions with the infinite size at the boundary. So, only the omega exchanges remain
at the large $N_c$. At the low baryon density, we have the repulsive skyrmion-omega-skyrmion interaction and
the positive energy density[33].
\section{Baryon Binding Energy}
In this section we are going to obtain the baryon binding energy from noncritical holographic model.
In previous section, we obtain the Lagrangian density of the baryon, so we write the energy of the
baryon as follows,
\begin{eqnarray}
E_{Baryon} &=& \int dx^3\int_{-\infty}^{+\infty} dZ\, ( \mathcal{L}_{Baryon})
\nonumber\\
&=&  2\,B\, V \int_{0}^{+\infty} dZ\,  K^{3/10}\sqrt{1 +
\frac{n_q^2}{ 4 B^2 \acute{B}}K^{-6/5}} ,
\end{eqnarray}
In order to obtain the baryon binding energy, in first step, we must subtract the self-energy of the $N_c$
quarks from this energy. Then by minimizing the result, we obtain the binding energy of the baryon.
The energy of the $N_c$ fundamental quarks in the noncritical background $AdS_6$ can be written as,
\begin{eqnarray}
S_{quarks} = \frac{N_c }{2 \pi \alpha'} \int \,dt\, dU  f(U)^{-1/2},
\end{eqnarray}
where we consider the condition $\acute{\tau}=0$.
\\Using the equations (11) and (12) in the above action, the energy of $N_c$ fundamental quarks is
written as follows,
\begin{eqnarray}
E_{quarks} = -C \int \, dZ \, Z \, K^{3/10}  ,
\end{eqnarray}
where
\begin{eqnarray}
C = \frac{2}{5}\frac{N_c \, U_{KK}}{2 \pi \acute{\alpha}}  .
\end{eqnarray}
So,  we can calculate the final energy as,
\begin{eqnarray}
E &=& E_{Baryon}-E_{quarks}
\nonumber\\
 &=&\,2\, V \int_{0}^{Z_{\Lambda}} dZ\,  K^{3/10}\sqrt{1 +
\frac{n_q^2}{ 4 B^2 \acute{B}}K^{-6/5}} \, -2\,C \int_{0}^{Z_{\Lambda}} \, dZ \, Z \, K^{3/10}
\end{eqnarray}
Minimizing this energy respect to the $Z_{\Lambda}$  results the baryon binding energy.
The equation (38) is solved numerically by the $N_f=2$, $N_c=3$, $M_{KK}=1GeV$ and $\acute{\alpha}=1$ values
and the baryon energy as a function of the $Z_{\Lambda}$ parameter has been shown in the Figure 4. As it is
clear from this figure, by increasing the $Z_{\Lambda}$, energy of baryon configuration decreases and has a
stable equilibrium point at $Z_{\Lambda}=Z_{min}$ with the minimum energy of $E_I = -5.003\, GeV$.
At $Z_{\Lambda} = Z_c$ the binding energy is zero again, and for $Z_{\Lambda}\,>\,Z_c$ the baryon would
be dissociated. So, we obtain an stable range for the baryon configuration. Also we can describe the
$ Z_{\Lambda}=Z_{min}$ point as the size of the baryon. Note that we obtained a similar behavior for the
 binding energy versus $Z_{\Lambda}$ compared to Ref.[14] with the $D4/D8-\overline{D8}$ configuration.
\\We show the values of baryon binding energy and the $Z_{min}$ for different values of $n$ in Table 1.
This Table indicated that for small $n$ values the binding energy is independent of $n$ value. It means
that in $(n_B/n_0)<1$, we can simplify the integrands to obtain the approximated energy as,
\begin{eqnarray}
\acute{E} =\,2\, V \int_{0}^{Z_{\Lambda}} dZ\,  K^{3/10}(1 +
\frac{(N_c\,n_0\,n)^2}{ 8 B^2 \acute{B}}K^{-6/5}+...) \, -2\,C \int_{0}^{Z_{\Lambda}} \, dZ \, Z \, K^{3/10}
\end{eqnarray}
which has analytical solution in terms of the following hypergeometric functions,
\begin{eqnarray}
\acute{E} &=&  \frac{0.4\,V}{B \acute{B}}Z_{\Lambda} [ 4.94\, B^2 \, \acute{B}\, F([-0.3,0.5],[1.5],-Z_{\Lambda}^2)
+0.62\,n_q^2 \, F([0.5,0.9],[1.5],-Z_{\Lambda}^2) ]
\nonumber\\
&-& 2 \,C\, Z_{\Lambda}\, F([0.3,0.5],[1.5],-Z_{\Lambda}^2),
\end{eqnarray}
\begin{figure}[tbp]
   \begin{center}
   \includegraphics[width=10cm]{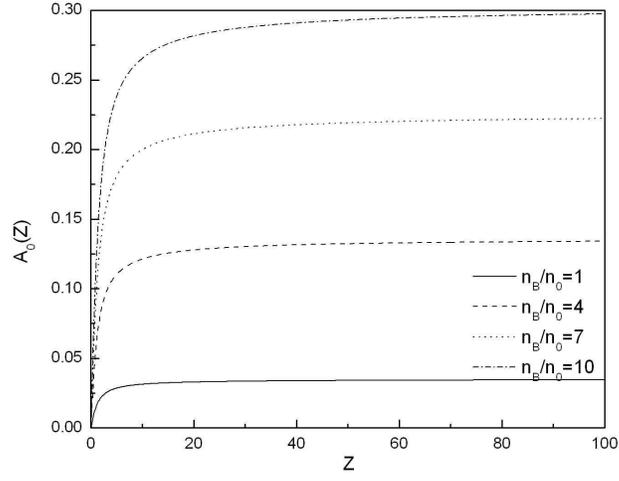}
      \caption{The gauge field $A_0(Z)$ vs Z.}
         \end{center}
\end{figure}
\begin{figure}[tbp]
\begin{center}
   \includegraphics[width=10cm]{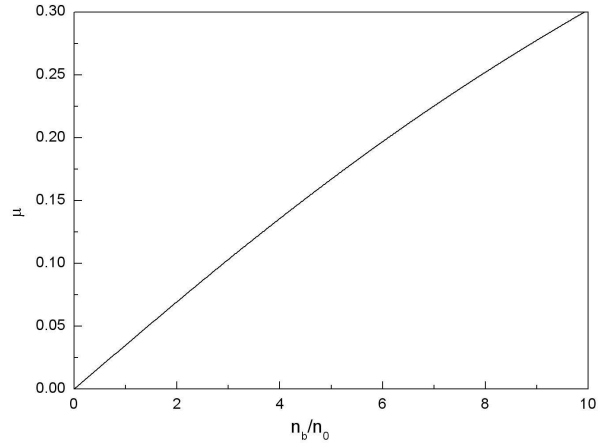}
   \caption{The chemical potential vs baryon charge $(n_B/n_0)$.}
    \end{center}
\end{figure}
\begin{figure}[tbp]
\begin{center}
   \includegraphics[width=16cm]{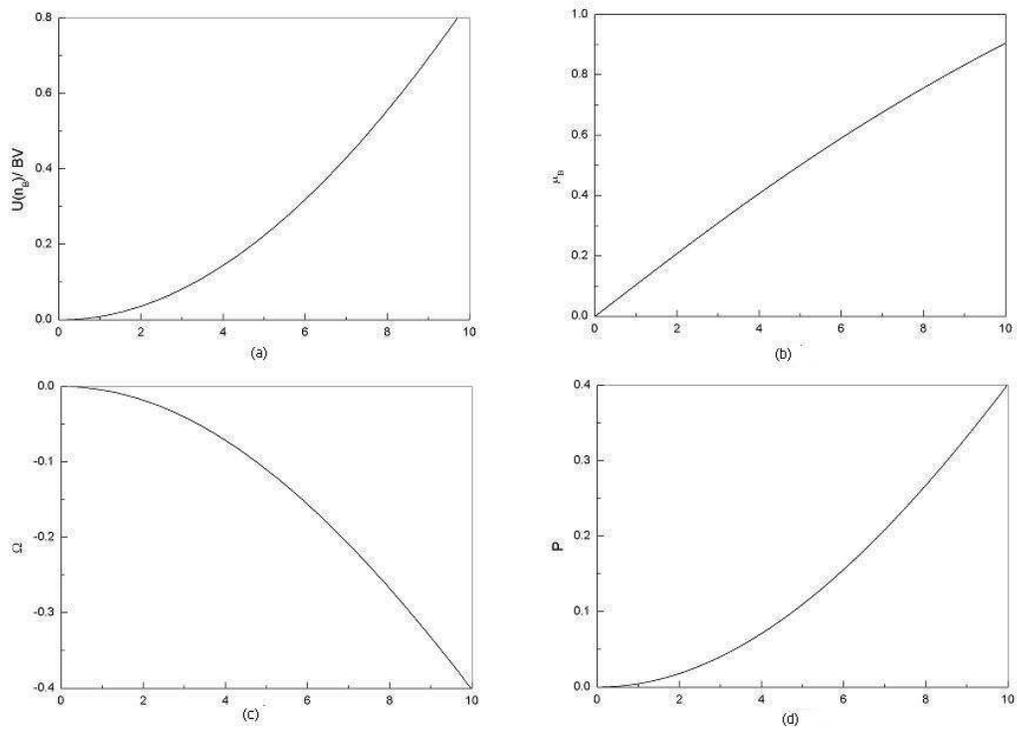}
   \caption{ (a)The internal energy,  (b) Baryon chemical potential,  (c) grand potential and (d)
 pressure vs baryon charge density $\frac{n_B}{n_0}$, where $n_B$ is the baryon density
and $n_0$ is the nuclear matter density.}
\end{center}
\end{figure}
\begin{figure}[tbp]
\begin{center}
   \includegraphics[width=10cm]{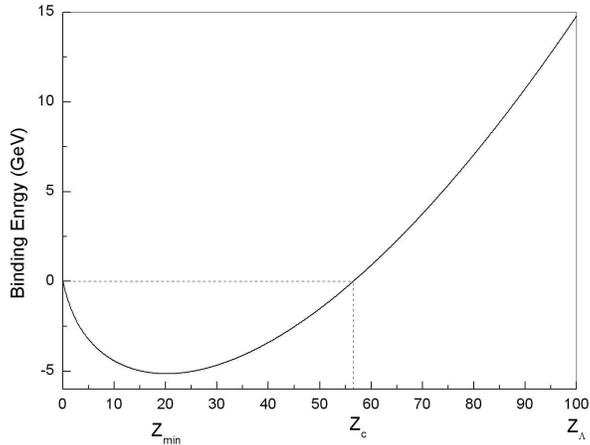}
   \caption{The behavior of baryon energy vs the cutoff parameter $Z_{\Lambda}$ The minimum of the energy
   shows the binding energy of the baryon and the $Z_{\Lambda}=Z_{min}$ can be regarded as the baryon size.}
\end{center}
\end{figure}
\begin{table}[htb]
\caption{\small The values of the $Z_{min}$ and the $E_I$ for the various baryon chemical potential. }
\begin{center}
\begin{tabular}{|c|c|c|c|c|c|}
\hline$\quad n\quad $   & $\quad10\quad$ & $\quad 7\quad$ & $\quad5\quad$ & $\quad1\quad$& $\quad10^{-3}\quad$ \\
\hline
$Z_{min}$  & 20.1 & 20.2&  20.2& 20.2& 20.2\\
\hline
$\quad E_I\quad$  & -5.11 &-5.12 &  -5.13&-5.14& -5.14\\
\hline
\end{tabular}
\end{center}
\end{table}
\section{Conclusion}
In this study, we considered the baryon in the noncritical
holographic model constructed by intersection
$D4/D4-\overline{D4}$ branes. This model have the $AdS_6$
background geometry. Here, we considered the baryon action as the
sum of the $DBI$ action and a delta function source of the gauge
field analogy to the SS model. We obtained the thermodynamics functions
in terms of the model parameters and studied the behavior of these
function respect to the baryon density. At low baryon densities, internal energy, grand
potential and the pressure are quadratic in $(n_B/n_0)$. But the
chemical potential has different behavior as indicated in equation (30). These
behaviors are similar to the results obtained from the SS model[33].
\\Also, we obtained the baryon binding energy in this model. In order to do this, we subtracted
the self-energy of the quarks from the total baryon energy. Then by minimizing the result
respect to the cutoff parameter, we obtained the binding energy for the baryon. We presented this
energy in the Figure 4. According to this Figure, we can easily find that the baryon energy
is zero at $Z_{\Lambda}=0$ or $U_{\Lambda}=U_{kk}$ which is the lower bound for $U$ coordinate in the model
where the radius of $S^{1}$ diminishes to zero and no stable baryon configuration exists. As $Z_{\Lambda}$
increases, the energy of baryon configuration gets smaller. At $Z_{\Lambda}=Z_{min}$ there is an stable
equilibrium point which corresponds to the size of baryon in the model. For $Z_{\Lambda} > Z_{min}$ the energy
increases and
also at $Z_{\Lambda} = Z_c$ the energy vanishes again. It reveals the fact that for $Z _{\Lambda}> Z_c$
there is no stable baryon configuration and the baryon would be dissociated. This behavior of baryon energy
is similar to the one we obtained in ref[14].

\end{document}